\documentclass[12pt]{article}
\usepackage{cjp}
\usepackage{epsf}

\begin{document}

\newcommand{\half}{\frac{1}{2}}
\newcommand{\be}{\begin{equation}}
\newcommand{\ee}{\end{equation}}
\newcommand{\bdm}{\begin{displaymath}}
\newcommand{\edm}{\end{displaymath}}
\newcommand{\<}{\langle}
\renewcommand{\>}{\rangle}
\newcommand{\Tr}{\mbox{Tr}}

\def\dirac{{\bf \rm D}\!\!\!\!/\,}
\def\wilson{{\bf \rm W}}
\def\ham{{\bf \rm H}}
\def\gauge{{\bf \rm G}}
\def\mbham{{\cal H}}
\def\bmat{{\bf \rm B}}
\def\cmat{{\bf \rm C}}

\newcommand{\ewxy}[2]{\setlength{\epsfxsize}{#2}\epsfbox[10 60 640 570]{#1}}

\newcommand{\pl}{{\rm pl}}
\newcommand{\csw}{{C_{SW}}}
\def\scro{{\cal O}}
\def\order{{\cal O}}
\def\su3{$SU(3)$}

%\footnote{Talk presented at the Chiral 99 workshop in 
%Taipei, Taiwan, Sept. 13-18, 1999.}, 

\title{
The Overlap-Dirac Operator: Topology and Chiral Symmetry Breaking
\vskip-3.7cm\hfill \vbox{\baselineskip 0cm\hbox{\rm\small JLAB-THY-00-01} 
\hbox{\rm\small FSU-SCRI-99-74}}\vskip2.8cm}

\author{Robert G. Edwards$^{(a)}$,
Urs M. Heller$^{(b)}$, Rajamani Narayanan$^{(c)}$}
\address{
{\it (a)} Jefferson Lab,
12000 Jefferson Avenue, 
Newport News, VA 23606, USA\\
{\it (b)} SCRI, Florida State University,
Tallahassee, FL 32306-4130, USA\\
{\it (c)} American Physical Society, 
One Research Road,
Ridge, NY 11961, USA}

\date{December 15, 1999}

\maketitle

\begin{abstract}
We review the spectral flow techniques for computing the index of the
overlap Dirac operator including results relevant for SUSY Yang-Mills theories.
We describe properties of the overlap Dirac operator,
and methods to implement it numerically. We use the results from the
spectral flow to illuminate the difficulties in numerical calculations
involving domain wall and overlap fermions.
\end{abstract}


\section{Overlap and domain wall Dirac operators}

In these proceedings, we review some basic properties of the overlap
Dirac operator and how its index can be computed by spectral flow techniques.
One of the side results is that for fermions in the adjoint
representation of $SU(N)$ we find evidence for fractional topological charge.
The presentation is pedagogical with the intent of illustrating the
origin of numerical difficulties in simulating overlap and domain wall
fermions. Recent results from our work using overlap fermions can be found in
references~\cite{EHN2,DEHN,EHKN2,dubna_proc}.
%We discuss the nature of the zero mode
%This same method can be used for computing the index of the domain
%wall Dirac operator

The massive overlap Dirac operator derived from the overlap
formalism~\cite{overlap} is
\be
D_{\rm ov}(\mu) = \frac{1}{2} \left[ 1 + \mu + (1-\mu) \gamma_5
 \epsilon(\ham_L(m)) \right]
\label{eq:D_ov}
\ee
where $\ham_L(m)$ is a lattice hermitian Dirac-like operator describing
a single fermion species with a large negative mass.  The mass $m$ is
a regulator parameter for the theory.  In this work, we use the
hermitian Wilson-Dirac operator $\ham_w(m) = \gamma_5 D_{\rm
Wilson}(-m)$, although we have tested other fermion actions.
The mass parameter $-1 < \mu <1$ is related to the fermion mass
by~\cite{EHN1} 
\be
m_f = Z_m^{-1} \mu (1 + {\cal O}(a^2)) .
\ee
The propagator for external fermions is given by
\be
{\tilde D}^{-1}(\mu) = (1-\mu)^{-1} \left[ D_{\rm ov}^{-1}(\mu) -1 \right] ,
\label{eq:prop}
\ee
{\it i.e.} it has a contact term subtracted, which makes the massless
propagator chiral: $\{ {\tilde D}^{-1}(0), \gamma_5 \} = 0$.

%For small momenta and $\mu$ the fermion propagator is related to
%the continuum propagator by
%\be
%D_c^{-1}(m_f) = Z_\psi^{-1} {\tilde D}^{-1}(\mu)
%\ee
%with the wave-function and mass renormalization satisfying $Z_\psi Z_m = 1$.

A massless vector gauge theory can also be obtained from domain wall
fermions~\cite{Kaplan}, where an extra, fifth dimension, of infinite extent
is introduced. 
In the version of ref.~\cite{Shamir},
one can show~\cite{HN3} that the physical (light) fermions contribute
$\log \det D_{\rm DW}$ to the effective action with the 4-d action
\be
D_{\rm DW} = \frac{1}{2} \left[ 1+\mu + (1-\mu) \gamma_5
\tanh \left( - \frac{L_s}{2} \log T \right) \right]
\label{eq:D_DW}
\ee
where $T$ is the transfer matrix in the extra dimension
and $L_s$ its size.
As long as $\log T \ne 0$ we obtain in the limit as $L_s \to \infty$
\be
D_{\rm DW} \rightarrow \frac{1}{2} \left[ 1+\mu + (1-\mu) \gamma_5
 \epsilon( - \log T ) \right].
\ee
This is just the massive overlap Dirac operator up to the replacement
$\ham_w \to - \log T$. It is easy to see that in the limit $a_s \to 0$,
where $a_s$ is the lattice spacing in the extra dimension (set to 1 above),
one obtains $- \log T = \ham_w \left( 1 + {\cal O}(a_s) \right)$.

\section{Some properties of the overlap Dirac operator}

In many cases it is more convenient to use the hermitian version of
the overlap Dirac operator (\ref{eq:D_ov}):
\be
H_o(\mu) = \gamma_5 D_{\rm ov}(\mu) = \frac{1}{2} \left[ (1+\mu) \gamma_5
 + (1-\mu) \epsilon(H_w)  \right] .
\label{eq:H_o}
\ee
The massless version satisfies,
\be
\{H_o(0), \gamma_5 \} = 2 H_o^2(0) .
\ee
It follows that $[H_o^2(0), \gamma_5] = 0$, {\it i.e.} the eigenvectors
of $H_o^2(0)$ can be chosen as chiral. Since
\be
H_o^2(\mu) = ( 1 - \mu^2 ) H_o^2(0) + \mu^2
\ee
this holds also for the massive case.

The only eigenvalues of $H_o(0)$ with chiral eigenvectors are 0 and $\pm 1$.
Each eigenvalue $0 < \lambda^2 < 1$ of $H_o^2(0)$ is then doubly degenerate
with opposite chirality eigenvectors. In this basis $H_o(\mu)$ and
$D_{\rm ov}(\mu)$ are block diagonal with $2 \times 2$ blocks, {\it e.g}
\be
%D_{\rm ov}(\mu) : \quad \begin{pmatrix}
%   (1-\mu) \lambda^2 + \mu & (1-\mu) \lambda \sqrt{1-\lambda^2} \\
%   -(1-\mu) \lambda \sqrt{1-\lambda^2} & (1-\mu) \lambda^2 + \mu
%   \end{pmatrix} ,
D_{\rm ov}(\mu) : \quad \pmatrix{
    (1-\mu) \lambda^2 + \mu & (1-\mu) \lambda \sqrt{1-\lambda^2} \cr
    -(1-\mu) \lambda \sqrt{1-\lambda^2} & (1-\mu) \lambda^2 + \mu
%    } ,
    } ,\qquad\gamma_5 = \pmatrix{ 1 & 0 \cr 0 & -1 } .
\ee
%where
%\be
%%\gamma_5 = \begin{pmatrix} 1 & 0 \\ 0 & -1 \end{pmatrix} .
%\gamma_5 = \pmatrix{ 1 & 0 \cr 0 & -1 } .
%\ee

\begin{figure}[t]
\begin{center}
\epsfysize 50mm
\centerline{\epsfbox[200 35 420 320]{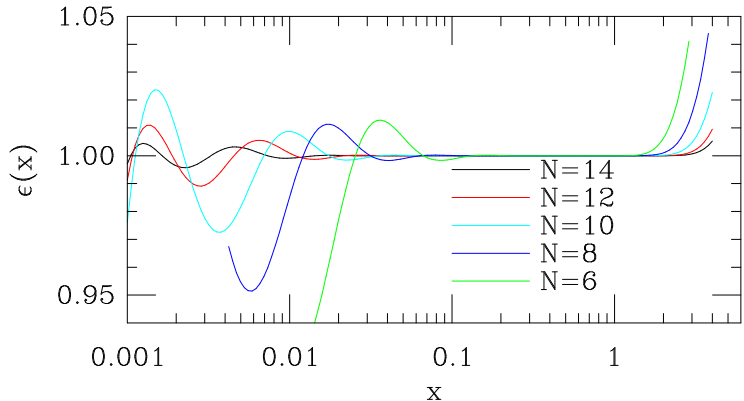}}
\figcaption{Plots of the optimal rational function approximation to
$\epsilon(x)$ for various order polynomials.}
\label{fig:Remez}
\end{center}
\end{figure}

For a gauge field with topological charge $Q \ne 0$, there are, in addition,
$|Q|$ exact zero modes with chirality ${\rm sign}(Q)$, paired with
eigenvectors of opposite chirality and eigenvalue 1. These are also
eigenvectors of $H_o(\mu)$ and $D_{\rm ov}(\mu)$:
\be
%D_{\rm ov}(\mu)_{\rm zero~sector} : \quad \begin{pmatrix}
%                         \mu & 0 \\ 0 & 1 \end{pmatrix}
D_{\rm ov}(\mu)_{\rm zero~sector} : \quad \pmatrix{
                         \mu & 0 \cr 0 & 1 }
  \qquad {\rm or} \qquad
%  \begin{pmatrix} 1 & 0 \\ 0 & \mu \end{pmatrix}
  \pmatrix{ 1 & 0 \cr 0 & \mu }
\ee
depending on the sign of $Q$.

We remark that from eigenvalues/vectors of $H_o^2(0)$ those of both
$H_o(\mu)$ and $D_{\rm ov}(\mu)$ are easily obtained. There is no need
for a non-hermitian eigenvalue/vector solver! For example, the Ritz
algorithm~\cite{ritz} will do just fine.

\begin{figure}[t]
\begin{center}
\centerline{{\setlength{\epsfxsize}{105mm}\epsfbox[120 250 520 400]{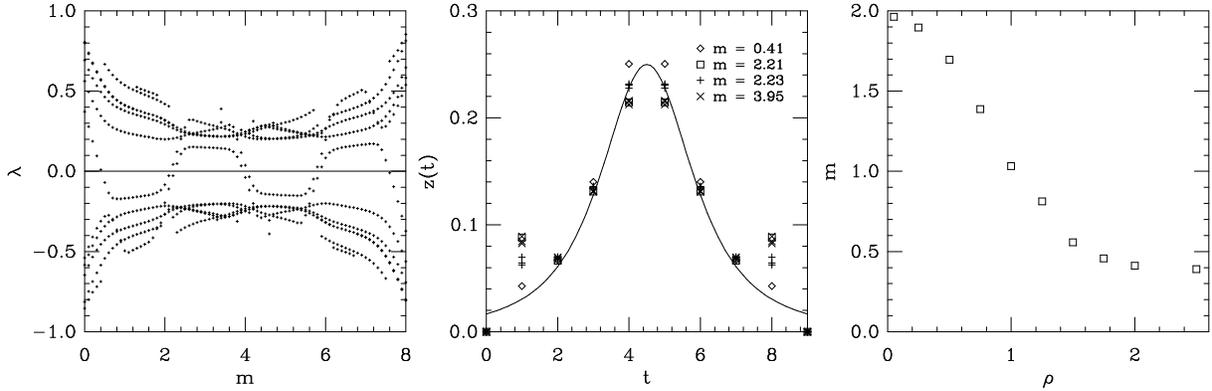}}}
\end{center}
\figcaption{Results for a single $8^4$ instanton with radius
$\rho=2.0$ and Dirichlet boundary conditions. 
Left: spectral flow
of $\ham_L(m)$, center: profile of the zero modes, right: mass
crossing value as a function of the instanton radius.}
\label{fig:wilson_inst}
\end{figure}

\begin{figure}[t]
\begin{center}
\centerline{{\setlength{\epsfxsize}{95mm}\epsfbox[120 220 460 420]{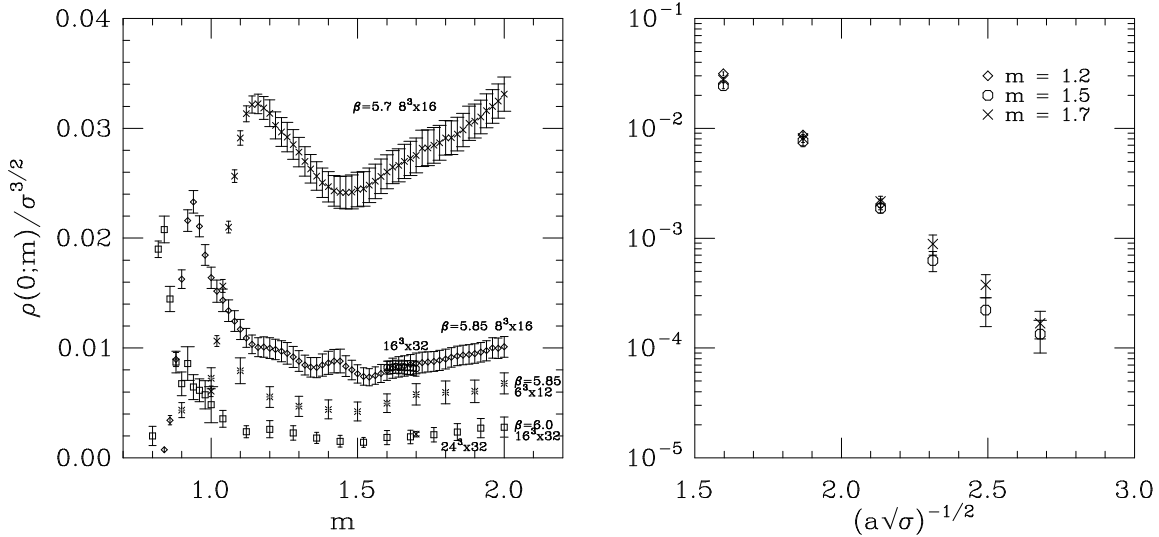}}}
\end{center}
\figcaption{On the left
$\rho(0;m)$ of $\ham_w(m)$ for quenched Wilson $\beta=5.7$, $5.85$ and $6.0$.
On the right, the approach of $\rho(0;m)$ to the continuum limit in the
quenched theory at fixed masses.}
\label{fig:rho_0}
\end{figure}

\begin{figure}[t]
\begin{center}
\centerline{{\setlength{\epsfxsize}{75mm}\epsfbox[50 120 570 550]{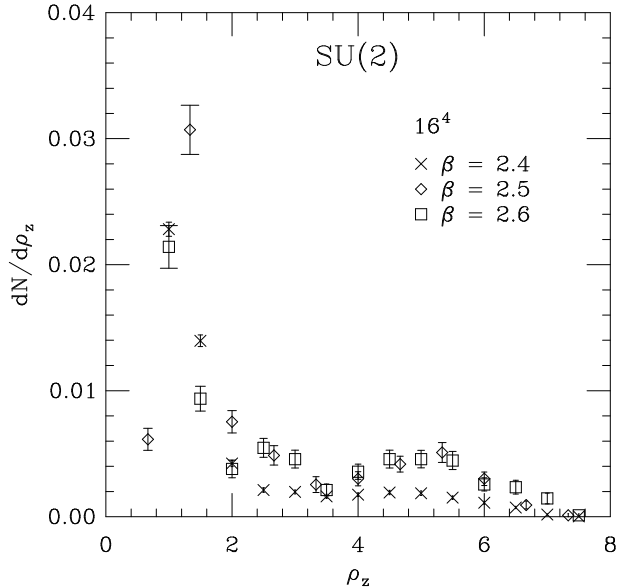}}}
\end{center}
\figcaption{Zero mode size distribution in lattice units for quenched
$16^4$ SU(2) Wilson gauge action.}
\label{fig:dn_su2}
\end{figure}

\begin{figure}[t]
\begin{center}
%\vspace{5mm}
\centerline{{\setlength{\epsfxsize}{70mm}\epsfbox[100 130 520 500]{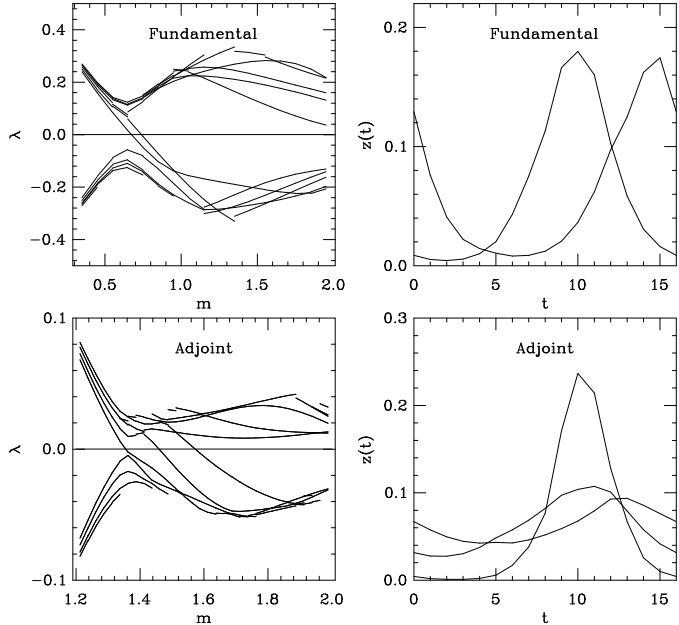}}}
\end{center}
\figcaption{Spectral flow and corresponding profile for SU(2)
configurations with $I_a \ne 4 I_f$ where $I_a$ is the number of
crossings in the adjoint rep. and $I_f$ is the number of crossings in
the fund. rep. There is a degeneracy of 2 in all crossings in the
adjoint rep.}
\label{fig:susy}
\end{figure}

\begin{figure}[t]
\begin{center}
%\vspace{5mm}
\centerline{{\setlength{\epsfxsize}{70mm}\epsfbox[50 110 570 500]{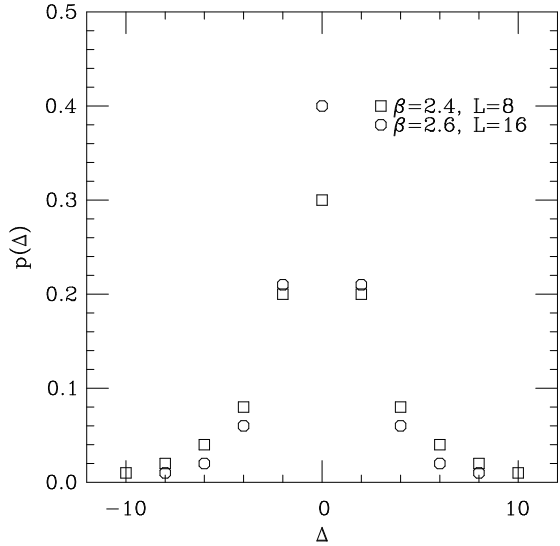}}}
\end{center}
\figcaption{Probability $p(\Delta)$ versus $\Delta$ for two gauge
ensembles where $\Delta=I_a-4I_f$.}
\label{fig:delta}
\end{figure}

\begin{figure}[t]
\begin{center}
\centerline{{\setlength{\epsfxsize}{70mm}\epsfbox[180 220 480 420]{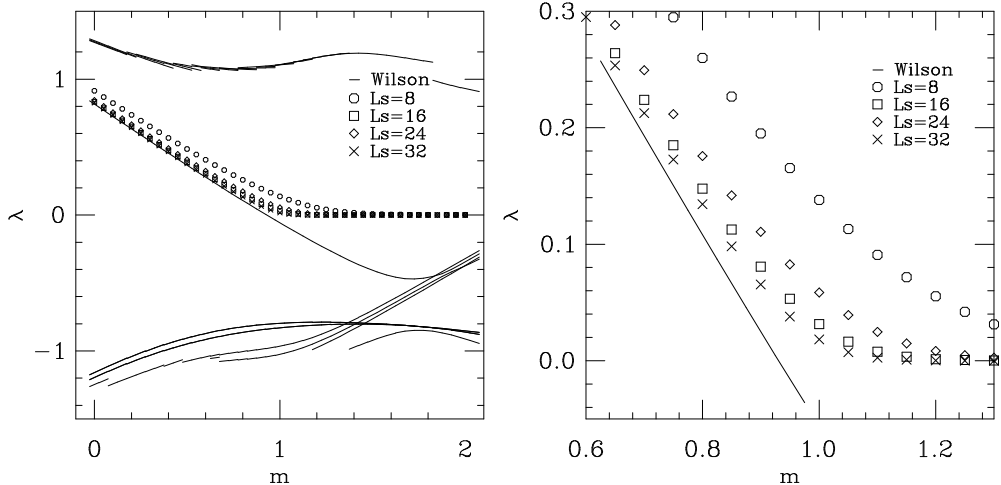}}}
\end{center}
\figcaption{Spectral flow of the hermitian domain wall Dirac operator $H_{DW}(m)$
and Wilson $H_w(m)$ on a single instanton background. Shown are 5D
extents of $L_s = 8$, $16$, $24$ and $32$. For $L_s=32$, only at a
mass separation of $0.2$ units from the crossing has the eigenvalue
dropped to $10^{-3}$.}
\label{fig:dwf_inst}
\end{figure}

\begin{figure}[t]
\begin{center}
\centerline{{\setlength{\epsfxsize}{95mm}\epsfbox[140 220 480 440]{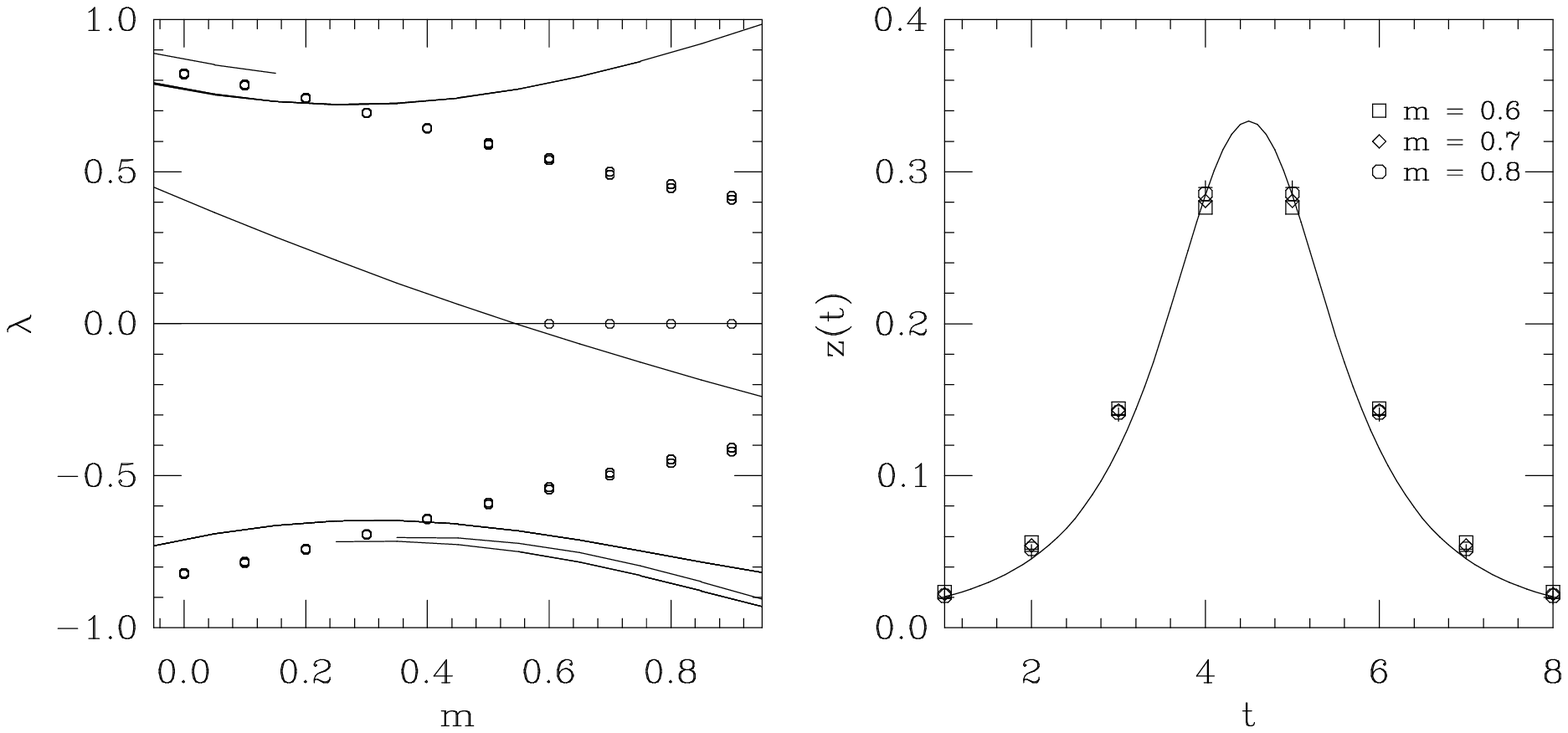}}}
\end{center}
\figcaption{Spectral flow of the overlap Dirac operator $H_o(m)$
and Wilson $H_w(m)$ on a single instanton background. On the right are
the profiles of the overlap zero modes compared to the t'Hooft solution.}
\label{fig:overlap_inst}
\end{figure}

\section{Implementations of the overlap Dirac operator}

In practice, we only need the application of $D(\mu)$ on a vector,
$D(\mu) \psi$, and therefore only the sign function applied to a
vector, $\epsilon(H_w) \psi$. Since we need the sign function of
an operator (a large sparse matrix) this is still a formidable task.
%\be
%D(\mu)\psi = \frac{1}{2} \left[1+\mu + (1-\mu) \gamma_5\epsilon(H_w)
% \right]\psi .
%\ee

Methods proposed for this computation are:
\begin{itemize}
\item A Chebyshev approximation of $\epsilon(x)=\frac{x}{\sqrt{x^2}}$
over some interval $[\delta,1]$~\cite{HJL}. For small $\delta$ a large
number of terms are needed.

\item A fractional inverse method using Gegenbauer polynomials for
$\frac{1}{\sqrt{x^2}}$~\cite{Bunk}. This has a poor convergence since
these polynomials are not optimal in the Krylov space.

\item Use a Lanczos based method to compute $\frac{1}{\sqrt{x^2}}$ based
on the sequence generated for the computation of
$\frac{1}{x}$~\cite{Borici}.

\item Use a rational polynomial approximation for $\epsilon(x)$ which
can then be rewritten as a sum over poles:
\be
\epsilon(x) \leftarrow x \frac{P(x^2)}{Q(x^2)} = 
  x \Bigl(c_0 + \sum_k \frac{c_k}{x^2 + b_k}\Bigr)
\ee
The application of $\chi\leftarrow\epsilon(H_w)\psi$ can be done by
the simultaneous solution of the shifted linear systems~\cite{Jegerlehner}
\be
(H_w^2 + b_k)\phi_k = \psi,\qquad \chi = H_w (c_0 \psi + \sum_k c_k \phi_k) .
\ee
One such approximation, based on the polar decomposition~\cite{pensacola},
was introduced in this context by Neuberger~\cite{HN2}.
We use optimal rational polynomials~\cite{EHN2}. The accuracy of this
approximation is shown in Fig.~\ref{fig:Remez}.
\end{itemize}

We note that in all methods listed above, one can enforce the accuracy
of the approximation of $\epsilon(x)$ for small $x$ by projecting out
the lowest few
eigenvectors of $H_w$ and adding their correct contribution exactly. 
%\begin{eqnarray*}
%\epsilon(H_w) &=& \sum_{i=1}^n |\psi_i\> \epsilon(\lambda_i) \<\psi_i|
% + {\cal P}_\perp^{(n)} {\rm App}[\epsilon(H_w)] {\cal P}_\perp^{(n)} , \\
%{\cal P}_\perp^{(n)} &=& {\bf 1} - \sum_{i=1}^n |\psi_i\> \<\psi_i| .
%\end{eqnarray*}
\be
\epsilon(H_w) = \sum_{i=1}^n |\psi_i\> \epsilon(\lambda_i) \<\psi_i|
 + {\cal P}_\perp^{(n)} {\rm App}[\epsilon(H_w)] {\cal P}_\perp^{(n)} ,
\qquad {\cal P}_\perp^{(n)} = {\bf 1} - \sum_{i=1}^n |\psi_i\> \<\psi_i| .
\label{eq:project}
\ee

To invert $D^\dagger D$ for overlap fermions, we have, generically, an outer
CG method (a 4-d Krylov space search) and an independent inner search method
for $\epsilon(H_w) \psi$ -- maybe CG again. For domain wall fermions, on the
other hand, a 5-d Krylov space search method is used.
It may pay off to try to combine inner and outer CGs for overlap
fermions by reformulating them into a 5-d problem~\cite{HN4,Borici2}.

%\begin{itemize}
%\item Modify the domain wall action so that the 4D overlap action is
%recovered after integration over extra flavors, such as
%$\tanh(\frac{L_s}{2}H_w)$ instead of
%$\tanh(-\frac{L_s}{2}\log(T))$.
%
%\item Rewrite rational approximation
%of $\epsilon(H_w)$ as a continued fraction. The solution of
%$D\psi=(1+\gamma_5 P(H)/Q(H))\psi=b$ can then be rewritten as a 5D
%operator, block-tridiagonal in the 5th direction.
%
%\end{itemize}

\section{Index defined via the Overlap formalism}

The massless overlap Dirac operator is
\be
D_{\rm ov} = \frac{1}{2} \left[ 1 + \gamma_5 \epsilon(\ham_L) \right];\qquad
\ham_L = \gamma_5 D_w(-m)\ .
\ee
%\begin{eqnarray}
%\ham^+_L = \gamma_5 D_w(-m)
%  = \pmatrix{ \bmat(U) -m & \cmat(U) \cr
%            \cmat^\dagger(U) &  -\bmat(U) +m }; && \!\!\!
%\ham^-_L = \gamma_5 = \lim_{m\rightarrow -\infty} \frac{1}{|m|}\ham_L^+(m).
%\end{eqnarray}
%\be
%\ham_L^+(m) = \gamma_5 D_{\rm Wilson}(-m) =
%%\begin{pmatrix} \bmat(U) -m & \cmat(U) \\
%%                \cmat^\dagger(U) &  -\bmat(U) +m \end{pmatrix}
%\pmatrix{ \bmat(U) -m & \cmat(U) \cr
%                \cmat^\dagger(U) &  -\bmat(U) +m }
%\ee
%\be
%\lim_{m\rightarrow -\infty} \frac{1}{|m|}\ham_L^+(m) = \gamma_5 = \ham^-
%\ee
%$C$ is the naive lattice Weyl operator, and $B$ is
%the laplacian. 
The index is given by $Q={\rm tr}\;\epsilon(\ham_L)/2$.
We see $Q$ simply counts the deficit of the number of positive
energy states of $\ham_L$. A simple method of computing
$Q$ at some fixed $m$ is via the spectral flow method
\cite{EHN2}.
Consider the eigenvalue problem
\be
\ham_L(m) \phi_k(m) = \lambda_k (m) \phi_k(m), \qquad
\frac{d\lambda_k(m)}{dm} = - \phi^\dagger_k(m) \gamma_5 \phi_k(m)\ .
\ee
An efficient way to compute $Q$ is to compute the lowest eigenvalues
of $\ham_L(m)$ for $m > 0$. We can prove the number of positive and
negative eigenvalues of $\ham_L(m)$ for $m < 0$ must
be the same, so we slowly vary the mass $m$ from $m=0$ while keeping
track of the levels crossing zero and direction of crossings up to
some $m$. In this way, we get the topological charge as a function of
$m$.

We note the mass $m$ must be greater than the usual critical mass of
$\ham_L(m)$, otherwise no topology change occurs and the overlap
operator does not describe a massless chiral fermion. This critical
mass value shifts from its free field value $0$ to some positive value
for non-zero gauge coupling. We should also choose a mass below $2$ so
that in the continuum limit there are no doubler contributions.

In Fig.~\ref{fig:wilson_inst}, we show spectral flow results for a
smooth background field of a single instanton~\cite{su2_inst_98}.
There is a reflection symmetry about $m=4$, namely the spectrum for
$8-m$ is opposite that of $m$. We see a mode crosses down in eigenvalue,
then crosses up again near $2$. There is a degeneracy of $3$ for the
modes just beyond $2$. Hence, as we increase $m$ we find a sequence of
the index $Q$ of $1$, $-3$, $3$, $-1$ and $0$ for $m=1$, $3$, $5$ and $8$.
%{\bf NOTE, MENTION DWF??.}
The one-dimensional profile for the modes associated with the crossings are
\be
z(t)=\sum_{\vec n} \phi^\dagger_k (\vec n, t)\phi_k (\vec n, t)
\label{eq:profile}
\ee
which we compare with the continuum solution in the center panel of
Fig.~\ref{fig:wilson_inst}
\be
z(t;c,\rho)= \frac{1}{ \Bigl[ 2\rho \bigl(
1 + (\frac{(t-c_4)}{\rho})^2 \bigr)^{3/2} \Bigr ] } \quad .
\label{eq:hooft_zero_mode}
\ee
Also shown is the zero crossing point $m$ as a function of $\rho$. We
see that as the size of the instanton decreases the crossing point
(the mass) increases.

As we turn on gauge fields, the picture of the flows complicates and
we can find crossings throughout the mass region beyond the critical
mass~\cite{su3_top_98,EHN3}. Since we are interested in the zero modes
at the crossings, we compute the density of zero eigenvalues
$\rho(0;m)$ of $\ham_w(m)$ by fitting linearly to the integrated density
\be
\int_0^\lambda \rho(\lambda') \,d\lambda' = 
  \rho(0)\lambda + {\frac{1}{2}}\rho_1\lambda^2 + \ldots
\ee
In Fig.~\ref{fig:rho_0} we show $\rho(0;m)$ for quenched SU(3)
lattices. We see that for $m$ beyond the critical mass region, the
density $\rho(0;m)$ rises sharply to a peak, then drops but is never
zero, hence the spectral gap is always closed. A similar result is
also found for two flavor dynamical fermion
backgrounds~\cite{su3_top_98} (simulated with positive physical quark
mass, {\em e.g.} not simulated in the super-critical mass region).
From a size distribution, we observe the zero modes are on the
order of one to two lattices spacings for $m$ beyond the main band of
crossings, namely for $m$ in the ``flat'' region of $\rho(0;m)$. 
We find that a physical quantity, like the topological susceptibility,
appears constant within errors in this ``flat'' region, indicating
that these small modes make no physical contribution.

For topology to change in a gauge field evolution, we must create
dislocations. These produce the small modes observed above which force
the spectral gap of $\ham_w(m)$ to be closed.  In the right panel of
Fig.~\ref{fig:rho_0} an empirical fit of the density to an exponential
of the inverse lattice spacing is shown. This result implies the
density is only zero in the continuum limit.  The density of zero
eigenvalues of $H_w(m)$, $\rho(0;m)$, is non-zero in the quenched
case, but decreases rapidly with decreasing coupling~\cite{EHN3}.
Very roughly, we find $\rho(0;m)/\sigma^{3/2} \ \sim \
e^{-e^\beta}$. We note the gauge action can be modified to reduce
$\rho(0;m)$, or even eliminate it altogether at some fixed
$m$~\cite{HJL2}.

In Fig.~\ref{fig:dn_su2}, we show the size distribution for quenched
$16^4$ SU(2) ensembles. The profile from Eq.~(\ref{eq:profile}) is
used to define a size
motivated by the t'Hooft zero mode in Eq.~(\ref{eq:hooft_zero_mode})
\be
\rho_z = {\frac{1}{2}}  {{\sum_t z(t)} \over {z_{\rm max}}}
\ee
We see there is a large number of modes about 1 to 2 lattice units
in size. There is a corresponding secondary peak around 5 lattice
units and all the distributions are bounded in size. This result
indicates that the size distribution of zero modes does not
show evidence for a peak at a physical scale (as suggested by some models)
even after we remove the small modes which are most likely lattice
artifacts~\cite{EHN3}. Instead, the observed scaling in lattice units
is suggestive of a finite volume effect.

\section{Evidence for fractional topological charge}

In a continuum background field with topological charge $Q$, the index
of the massless Dirac operator in the {\it adjoint} representation is
equal to $2NQ$~\cite{ad_index,su2_adj_98}.  Classical instantons carry
integer topological charge and can thus only cause condensation of an
operator with $2N$ Majorana fermions.  Witten argued that a {\it
bilinear} gluino condensate exists in SUSY YM.  Self-dual twisted
gauge field configurations, with fractional topological charge $1/N$
exist (t'Hooft), and could explain a bilinear gluino condensate.  What
about non-classical gauge field configurations?

The adjoint representation is real $\Rightarrow$ spectrum of
$\ham_L$ is doubly degenerate: adjoint index can only be even.  Are
all even values realized, or only multiples of $2N$ ? To this end, we
studied the flow in the adjoint representation on two SU(2) ensembles
at the same fixed physical volume~\cite{su2_adj_98}. We do find
configurations with $I_a = 4 I_f$ (number of adjoint and fundamental
crossings), but we also find configurations with $I_a \ne 4 I_f$. An
example is shown in Fig.~\ref{fig:susy}. 

To check if this evidence for fractional topological charge is a
lattice artifact, we plot $\Delta = I_a - 4I_f$ in
Fig.~\ref{fig:delta}.  Note that $\Delta$ takes on only even values.
The probability of finding a certain value of $\Delta$, $p(\Delta)$,
is plotted for two ensembles in Fig.~\ref{fig:delta}. We find that
$p(\Delta)$ for $|\Delta| > 2$ decreases as one goes toward the
continuum limit at a fixed physical volume.  However, $p(\pm 2)$ does
not decrease indicating that it might remain finite in the continuum
limit.

\section{Main problem for Overlap and Domain Wall fermions}

The existence of small eigenvalues illustrated in Fig.~\ref{fig:rho_0}
hampers the approximation accuracy and convergence properties of
implementations of $\epsilon(H_w)$. Eigenvector projection both
increases the accuracy of the approximation and decreases the
condition number, {\em e.g.} of the inner CG.

The existence of small eigenvalues has implications also for domain
wall fermions. One can show that the spectrum of $-\log{T(m)}$ of
Eq.~(\ref{eq:D_DW}) around zero is the same as the spectrum of
$H_w(m)$~\cite{overlap}. While the small eigenvalues of $-\log{T(m)}$
don't appear to cause algorithmic problems for domain wall fermions,
they can induce rather strong $L_s$ dependence of physical quantities,
and hence causing the need for large $L_s$.

\subsection{Domain Wall and Overlap-Dirac operator spectral flow for 
smooth SU(2)}

As an illustration of the effects of low lying modes of $H_w(m)$, we
show in Fig.~\ref{fig:dwf_inst} the spectral flow of the hermitian 
domain wall operator $H_{\rm DW}(m) = \Gamma_5 D_{\rm
DW}(m)$ on a smooth $4^2\times 8^2$ single instanton background with
Dirichlet boundary conditions (BC). The $\Gamma_5$ includes a parity
operator. Also shown is the corresponding hermitian Wilson $H_w(m)$
spectral flow. 
%There is a crossing indicating a net index of 1. 
The ``zero'' DWF eigenvalue sets in slowly in $L_s$. We need $L_s \sim
1/\lambda_{\rm min}$ where $\lambda_{\rm min}$ is the lowest eigenvalue of
$H_w(m)$ (which is similar to the lowest eigenvalue of $\log(T(m))$)
for $\epsilon(-\log{T(m)})\approx\tanh\left(-\frac{L_s}{2}\log{T(m)}\right)$.
In fact, the (almost) chiral zero mode eigenvalue 
$\lambda_{\rm DW}(m)\sim {\rm const}\times(1-\tanh(\lambda_{\rm min}(m)L_s/2))$ for
$m$ beyond the crossing indicating a sensitivity to the hermitian
Wilson operator eigenvalue.

In Fig.~\ref{fig:overlap_inst}, we show a similar plot of the
hermitian overlap Dirac operator $H_o(m)$ on a smooth
$8^4$ single instanton background with Dirichlet BC. %boundary conditions.
There are strict zero modes after the crossing, $m=0.6$, $0.7$, and
$0.8$. Also shown are the zero mode profiles for these masses which
are quite similar and nicely follow the t'Hooft zero mode solution.

\section{ Conclusions}

The overlap Dirac operator has the same chiral symmetries as
continuum fermions. It has exact zero modes in topologically
non-trivial gauge fields. It is therefore well suited for a study
of the interplay of topology, with its
associated exact zero modes, and chiral symmetry breaking,
determined by the density of small eigenvalues. 

The creation of dislocations necessary for change of topology causes
numerical difficulties with overlap and domain wall simulations. These
dislocations are purely a property of the gauge actions used and are
not a fundamental limitation of the chiral fermion formalisms. In
practice, the projection technique used in the overlap simulations is
vital to precisely control the adverse influence of the low lying zero
modes of the hermitian Wilson operator. The same technique can be used
for domain wall simulations, but it is more cumbersome. Further work
is directed towards lowering (and possibly eliminating) the density
of low lying zero modes. A dynamical HMC algorithm for the overlap
Dirac operator has also been developed~\cite{dubna_proc}.

%\section*{Acknowledgements}
RGE would like to thank the organizers for a splendid workshop.
The work of RGE and UMH has been supported in part by DOE contracts
DE-FG05-85ER250000 and DE-FG05-96ER40979.
RGE was also supported by DOE
contract DE-AC05-84ER40150 under which 
%the Southeastern Universities
%Research Association (SURA) 
SURA
operates the Thomas Jefferson National
Accelerator Facility.

\end{document}